\documentclass[compsoc,conference,a4paper,10pt,times]{IEEEtran}
\usepackage{url,amsmath,amssymb,amsfonts,booktabs,multirow}
\usepackage{graphicx}
\usepackage{tikz}
\usetikzlibrary{arrows.meta,positioning,calc}
\usepackage{array}
\usepackage{algorithm}
\usepackage{algpseudocode}
\usepackage{amsmath}

\begin{document}
\title{Static Attribution of Android Residential Proxy Malware Using Graph Kernels}
\author{
\IEEEauthorblockN{Peter Clark}
\IEEEauthorblockA{Iowa State University\\
Sandia National Laboratories\\
pgclark@\{iastate.edu,sandia.gov\}}
\and
\IEEEauthorblockN{Yong Guan}
\IEEEauthorblockA{Iowa State University\\
guan@iastate.edu}
\and
\IEEEauthorblockN{Zhonghao Liao}
\IEEEauthorblockA{Milwaukee School of Engineering\\
liao@msoe.edu}
}
\maketitle
\begin{abstract}
Android residential proxy applications represent a growing class of potentially unwanted programs (PUPs) that route third-party traffic through end-user devices, often covertly, providing infrastructure that threat actors use for ad fraud, credential abuse, and evasion of geolocation controls. Attributing an unknown APK to a specific proxy network remains challenging due to code reuse, SDK embedding, and obfuscation across proxy families.

We present a static-analysis pipeline for automated proxyware family attribution, extracting graph-structured representations (control-flow and function-call graphs) from a validated corpus of 3,781 Android proxy apps spanning eight commercial proxy networks. We evaluate Weisfeiler-Lehman graph kernel features against Drebin's code features, MaMaDroid's call-chain features and the LibScan library detector across multiple classifiers. Using 5-fold DEX-grouped cross-validation to prevent data leakage, a linear model on WL features achieves a macro F1 of 0.950, on par with code-only Drebin features (0.951). Grouping applications by host lineage, so that builds of the same host application are never split between training and test, lowers both, to 0.722 and 0.786. Against 4,887 independently collected Google Play applications, the WL model reaches an AUROC of 0.982, but at the threshold that retains 95\% of known proxy applications it still assigns a provider to 5.2\% of them.

Beyond attribution, we examine the persistence of proxy functionality in applications from our expanded corpus. Of the 1,508 corpus packages, 80\% are no longer listed on APKPure; of the 212 still analyzable, 65.6\% retain their original proxy-family label, and the provider's SDK markers confirm 93\% of those predictions. Further analysis of developer accounts reveals 23 developers whose other applications include 26 confirmed to carry a proxy SDK, suggesting that proxy SDK integration extends beyond the applications in our original corpus; on their catalogs the classifier is precise except for one template publisher, whose shared code it mistakes for the SDK. Together, these findings show that static proxy-family attribution is feasible, while highlighting the difficulty of generalizing across application lineages and of distinguishing known proxy families from unrelated applications.

\end{abstract}
\section{Introduction}
Residential proxies are a class of commercial proxy services that route customer traffic through end-user devices using IP addresses assigned by consumer Internet service providers. Unlike datacenter proxies, which use IP ranges that can often be associated with particular hosting providers, residential proxies make proxied traffic difficult to distinguish from ordinary browsing. However, this capability often comes at the expense of the device owner, as the proxy software is frequently deployed without meaningful consent through Android applications that embed third-party SDKs in exchange for monetization. These applications share characteristics with trojans in that they are distributed through mainstream app stores and rarely disclose their full network capabilities to users. \\
Measurement research by Mi et al.\ demonstrated that a substantial portion of residential proxy egress is facilitated through Android applications, and assembled a labeled dataset of 1,701 APKs whose network traffic connected to backconnect infrastructure of four major residential proxy providers: IPNinja, Luminati (now Bright Data), Monkeysocks, and Oxylabs. Their findings showed that consent dialogs for these apps were often misleading or absent entirely, and that proxied traffic through these apps was used for ad fraud and other criminal activity \cite{mi2021your}. Earlier work established that the majority of proxy networks other than Luminati lack transparent deployment pipelines, that these networks provide millions of IP addresses for customer use, and that the bulk of these addresses are not tracked by known blocklists \cite{mi2019resident}. \\
These services have direct operational relevance to sophisticated threat actors. In 2024, Midnight Blizzard, a Russian threat actor tied to the SVR, used residential proxies to compromise a legacy, non-production test tenant at Microsoft through password spraying, subsequently using OAuth applications to access corporate email accounts \cite{microsoft2024midnight,ncsc2024svr}. Scattered Spider has used commercial VPNs and residential proxy services, including Luminati and Oxylabs, for credential-stuffing, phishing, and social engineering campaigns in order to evade geolocation controls \cite{thomas2025scattered}. More recently, Google's Threat Intelligence Group disrupted the IPIDEA proxy network after observing over 550 threat actors using the network within a 7-day period. Through static and dynamic analysis of SDKs embedded in IPIDEA-affiliated applications, they reverse engineered a two-tier command-and-control architecture. By decoding the telemetry protocol, they were able to dynamically enumerate approximately 7,400 Tier 2 nodes and discovered that the same backconnect infrastructure was shared by several nominally independent VPN providers. Google identified over 600 Android applications containing code connected to IPIDEA's command-and-control infrastructure \cite{mandiant2026ipidea}. \\
Despite this growing operational relevance, academic work on residential proxy applications has remained largely within the domain of Internet measurement, leaving the analysis of the applications themselves relatively unexplored. Residential proxy functionality is commonly distributed as reused SDK code embedded within otherwise unrelated host applications, which gives static attribution an opening: applications of the same proxy family may share recognizable code structures even when their host functionality differs. The ability to attribute an unknown application to a specific proxy family has practical implications for threat intelligence, incident response, and the development of detection signatures that can identify proxy enrollment before a device begins routing third-party traffic.

We investigate two related tasks: attribution, the task of determining which family a known proxy app belongs to, and detection, the task of determining whether an arbitrary app contains proxy code. We evaluate attribution by macro-F1 score on the labeled corpus, and detection by applying the best-performing model to independently collected Google Play applications and to current APKPure versions and developer catalogs outside the corpus, which are previously unlabeled inputs.

\subsection{Contributions}
To our knowledge, this paper presents the first systematic evaluation of graph-based structural features for residential proxy family attribution. While prior work has characterized the network behavior, infrastructure, and abuse patterns of residential proxy services, we are unaware of academic work applying code-level structural analysis to the proxy applications themselves. The closest methodological precedents are Android third-party library detection systems, including LibScout~\cite{libscout}, LibID~\cite{libid}, LibScan~\cite{libscan}, and LibD~\cite{libd}, and graph-similarity approaches, including SAD~\cite{sadcdg}, MatchScope~\cite{matchscope}, and LibHunter~\cite{libhunter24}; these address library identification or application similarity rather than proxy family attribution, and we compare against LibScan directly (Section~\ref{sec:res_baselines}). The approach relies on residential proxy functionality being distributed as reusable SDK code, so that each proxy family can leave a structural fingerprint in its APKs; graph-based features detect these fingerprints with a macro-F1 of 0.950 over eight providers under DEX-grouped evaluation, and 0.722 when applications of the same host lineage are kept together (Table~\ref{tab:baselines}). Our contributions are as follows:
\begin{itemize}
    \item \textbf{Dataset expansion via DEX reuse:} We leverage DEX file hash overlap to systematically expand the Mi et al.\ proxy corpus from 1,629 to 3,365 labeled APKs, demonstrating a reproducible technique for growing labeled corpora in SDK-driven malware ecosystems, and show that labels propagated this way must be validated: 675 of the pivot-discovered applications are linked to their provider only through a library DEX file. The validated corpus, extended with version builds and four providers found through published command-and-control domains, holds 3,781 applications of eight providers.
    \item \textbf{Graph-based structural attribution:} We extract control-flow graphs (CFGs) and function-call graphs (FCGs) from the corpus and generate feature vectors using Weisfeiler-Lehman graph kernels. This represents the first structural code-level analysis of these applications in the academic literature, which has previously studied them only through network measurement and string matching.
    \item \textbf{Multi-classifier evaluation:} We evaluate linear and Random Forest classifiers using 5-fold DEX-grouped cross-validation to prevent data leakage, and under a stricter grouping by host lineage, achieving a macro F1 of 0.950 with a linear model on the eight-class attribution task. We compare WL features against Drebin's code features, MaMaDroid's call-chain features and the LibScan library detector: code-only Drebin features match WL under DEX grouping (0.951) and degrade less under lineage grouping (0.786 against 0.722), WL refinement adds nothing measurable over the initial node labels, and LibScan makes no incorrect attributions but detects a reference in only 52\% of applications.
    \item \textbf{Open-world evaluation:} We extend from an eight-class attribution setting to a nine-class open-world setting by adding a \texttt{non\_proxy} class containing 996 arbitrary applications, reaching a macro-F1 score of 0.941 across the 4,773 applications with extracted graphs. Against 4,887 independently collected Google Play applications, rejection by maximum class probability reaches an AUROC of 0.982, but at the threshold that retains 95\% of known proxy applications 5.2\% of Play applications are still assigned a provider, so classifier output is a triage signal rather than an attribution.
    \item \textbf{Measured limits:} Identifier renaming changes 70\% of the predictions of a WL model trained on unmodified applications, while a renaming-invariant coloring is unaffected at a cost of 4 to 7 points of macro-F1. A detector trained without a provider flags 0.5\% to 53\% of its applications at a 1\% false-positive rate, and a model trained on Luminati's current SDK recognizes at most 43\% of its natively implemented predecessor.
    \item \textbf{Ecosystem persistence:} Of the 1,508 corpus packages, 80\% are no longer listed on APKPure; of the 212 still analyzable, 65.6\% retain their original proxy-family label, 93\% of them confirmed by the provider's SDK markers. Across the catalogs of 23 developer accounts, verified precision is 0.93 except for one template publisher, on whose catalog the classifier recognizes shared template code rather than the SDK, so classifier output is a triage screen rather than an attribution.
\end{itemize}

\section{Background and Related Work}\label{sec:background}
Here we discuss previous work in the measurement of residential proxies, graph kernels, mobile malware analysis, and code reuse as they relate to the problem of proxyware family attribution.

\subsection{Residential proxies ecosystem}
A proxy is a technology for mediating traffic between a client and a server by intercepting traffic, breaking the connection at a certain layer of the network stack, and re-transmitting the data as originating from the proxy egress node. For example, a proxy that re-encapsulates a client's packets adds per-packet overhead, which can reduce the effective MTU observable at the network layer \cite{smutz2025taxonomy}. Residential proxies extend this model by routing customer traffic through egress points located on end-user devices whose IP addresses exist in address space allocated to residential consumer Internet service providers (ISPs). This allows their customers to forward traffic through egress points which more closely resemble benign usage patterns. In practice, there are often significant differences between TCP and TLS round-trip times sufficient to identify residential proxy traffic, suggesting that many operate as transport-layer proxies \cite{chiapponi2023poster}. \\
Most commercial residential IP proxies function via backconnect architecture where traffic is routed through a provider-internal network before reaching its egress point. This contrasts with direct-mode architectures, particularly prevalent in China, which directly expose the exit node to consumers and can be enumerated via techniques such as passive DNS \cite{choi2020understanding}. Hanzawa et al.\ analyzed communication patterns of residential proxy hosts in Japan, finding that most were mobile devices engaged in port scanning as their most frequent malicious activity \cite{hanzawa2020analysis}, while Chiapponi et al.\ demonstrated that web scraping bots increasingly leverage residential IP proxy services to defeat commercial bot countermeasures and proposed a detection method based on network measurements \cite{chiapponi2022badpass}.

A substantial body of subsequent work has studied the residential proxy ecosystem, but primarily from the perspective of network measurement and abuse characterization rather than analysis of the proxy applications themselves. Yang et al.\ conducted an extensive measurement study of residential proxies in China, characterizing their infrastructure and abuse patterns \cite{yang2022extensive}. Chiapponi et al.\ examined the industrial impact of web scraping through residential proxies \cite{chiapponi2022industrial} and later reported longitudinal measurement findings from large-scale proxy detection campaigns \cite{chiapponi2023inside}. Benhabbour and Dacier surveyed network disruption techniques including residential proxy abuse as a middlebox phenomenon \cite{benhabbour2025endemic}. Kikuchi et al.\ deployed honeypot-based residential proxies to observe malicious activities routed through them \cite{kikuchi2025honeyproxy}, and Huang et al.\ developed traffic classification techniques for identifying residential proxy flows at the network level \cite{huang2024shininglight}. Wang et al.\ audited the security and privacy practices of mobile VPN applications, some of which overlap with the residential proxy ecosystem \cite{wang2026mvpnalyzer}. 

The original dataset assembled by Mi et al.\ \cite{mi2021your}, which forms the basis of our corpus, used backconnect domain enumeration and network-level indicators to identify and label proxy applications, but their static analysis of the applications was limited to string matching. To the best of our knowledge, no prior academic work has applied structural code analysis or graph-based machine learning techniques to the residential proxy application corpus itself. Our work addresses this gap by shifting the analysis from network traffic and infrastructure to the internal structure of the applications.

\subsection{Graph kernels}\label{sec:graph_kernels}
Graphs are natural representations for many types of structured data, but their variable size and topology make them incompatible with standard machine learning algorithms that operate on fixed-dimensional feature vectors. A graph kernel addresses this by defining a positive semi-definite function $k : \mathcal{G} \times \mathcal{G} \to \mathbb{R}$ that measures similarity between pairs of graphs. By the Moore-Aronszajn theorem, any such function implicitly defines a reproducing kernel Hilbert space (RKHS) $\mathcal{H}$ and a feature map $\phi : \mathcal{G} \to \mathcal{H}$ such that $k(G_1, G_2) = \langle \phi(G_1), \phi(G_2) \rangle_{\mathcal{H}}$. The RKHS is a (possibly infinite-dimensional) vector space in which each graph is represented as a point allowing inner product $\langle \phi(G_1), \phi(G_2) \rangle_{\mathcal{H}}$ to quantify structural similarity between two graphs without requiring that the feature map $\phi$ be computed explicitly. This technique allows algorithms that depend only on inner products between data points, such as support vector machines, to operate in $\mathcal{H}$ by evaluating $k$ directly, avoiding the cost of constructing explicit high-dimensional representations \cite{vishwanathan2010graph}.

In practice, most graph kernels work by decomposing each graph into a bag of substructures , such as random walks, shortest paths, or rooted subtrees, and defining the kernel as the inner product of the resulting substructure count vectors. The choice of substructure determines what aspects of graph topology the kernel captures and at what computational cost \cite{vishwanathan2010graph}.

Among the most widely used graph kernels is the Weisfeiler-Lehman (WL) subtree kernel, introduced by Shervashidze et al.\ \cite{shervashidze2011weisfeiler}. The WL kernel is based on the 1-dimensional Weisfeiler-Lehman graph isomorphism test, an iterative procedure that refines node labels by aggregating neighborhood information. At each refinement iteration $h = 0, 1, \ldots, H-1$, every node $v$ receives a new label computed as a hash of its current label concatenated with the sorted multiset of its neighbors' labels:
\begin{equation}
\ell^{(h+1)}(v) = \text{\textsc{Hash}}\!\left(\ell^{(h)}(v),\; \{\!\!\{ \ell^{(h)}(u) : u \in \mathcal{N}(v) \}\!\!\}\right)
\end{equation}
where $\ell^{(h)}(v)$ is the label of node $v$ at iteration $h$, $\mathcal{N}(v)$ denotes the neighbors of $v$, and $\{\!\!\{ \cdot \}\!\!\}$ denotes a multiset. At each iteration, the multiset of all node labels across the graph defines a feature vector; concatenating these vectors over levels $h = 0, 1, \ldots, H$ produces the final WL feature representation. The kernel between two graphs is then the inner product of their respective feature vectors. The WL kernel admits an explicit finite-dimensional feature map: each unique label produced across all iterations corresponds to a dimension, and the feature vector counts occurrences of each label. This means the kernel can be evaluated either implicitly via the kernel function or explicitly by constructing feature vectors and computing their dot product, the latter being the approach we adopt in this work. Our implementation approximates this map with fixed-dimensional signed feature hashing and aggregates predecessor and successor labels separately to preserve edge direction in program graphs (Section~\ref{sec:methodology}).

The WL kernel captures local structural patterns at increasing neighborhood depths, scales approximately linearly in the number of edges for a fixed number of iterations (i.e., $O(Hm)$ for $m$ edges and $H$ iterations, given constant-time label hashing), and distinguishes many, though not all, non-isomorphic graphs \cite{shervashidze2011weisfeiler}. These properties make it well-suited to the analysis of program graphs, where structural patterns at different granularities, from individual basic blocks to multi-function call chains, may carry discriminative information; whether refinement adds information beyond the initial node labels is an empirical question that we test (Section~\ref{sec:res_baselines}).

\subsection{Graph-based malware classification}
Graph-based analysis of malware arose from the need to examine underlying structural features rather than surface-level indicators. Early work applied unsupervised clustering and edge-edit metrics to function-call graphs in order to support approximate matching, showing that malware families often reuse structural features if not code itself. Similar techniques have been applied to control-flow graphs, though these features are more sensitive to changes such as control-flow obfuscation \cite{bai2019malware}. \\
Supervised and unsupervised malware analysis has since trended towards learning representations of graph features in vector spaces, with a focus on subgraph mining. Work on graphlet sampling has examined distributions of key graphlets and their indicative value for malware commonality \cite{gao2018android}. Recent approaches now combine topological features with semantic indicators such as API sources and sinks or API call semantics, propagated along nodes and edges, in order to combine local behavioral contexts with larger-scale structural patterns \cite{lu2020afcgdroid,yang2021android}. These approaches have been found comparatively resilient when evaluated against real-world obfuscation and dataset scaling \cite{bilot2024survey}. \\
Among graph kernel methods, the WL kernel has seen particular adoption for malware classification due to its ability to efficiently capture subtree patterns while scaling linearly in the number of graph edges. Early work demonstrated the feasibility of WL kernels on Android control-flow graphs \cite{sahs2012machine}, and subsequent approaches have enhanced the kernel with contextual information from function call graphs \cite{ling2019android} and applied it to dynamic execution representations for packer identification \cite{li2019consistently}. WL-based kernels have been applied across multiple graph classification settings, owing to their simplicity and discriminative ability \cite{dalton2020classifying,chen2023guided}. Related lines use structural graph kernels for detection \cite{gascon2013structural}, contextual WL variants carrying semantic information \cite{narayanan2016cwlk}, and multiple-kernel learning that also localizes malicious code \cite{narayanan2018mkldroid}.

\subsubsection{Mobile malware analysis}
Early work in Android malware analysis focused on categorizing malware families based on characterizing sequences of API calls \cite{aafer2013droidapiminer}, though these approaches were found to be vulnerable to obfuscation and concept drift as applications evolve across versions \cite{pan2020systematic}. This led to the adoption of graph-based representations extracted from Dalvik bytecode, which capture structural invariants that persist across some surface-level code transformations, although identifier renaming, dead code insertion, and control-flow flattening can each alter them (Section~\ref{sec:robustness}). Recent work has also explored features beyond Dalvik bytecode, incorporating native code and multi-modal analysis to improve detection coverage \cite{sun2021android}. Graph-based techniques combined with node annotations such as sensitive API sources and sinks achieve higher F1 than structural features alone, allowing classifiers to jointly reason about what an application does and how its code is organized \cite{shi2023sfcgdroid}.

\subsection{Code reuse and library detection}
Code reuse is pervasive in Android software, where applications frequently incorporate third-party libraries and shared SDKs that complicate similarity-based malware analysis \cite{gonzalez2016measuring}. Zhang et al.~\cite{zhang2020pml} conducted a large-scale empirical study of potentially malicious libraries in over 500K Android apps, documenting how trojanized SDKs are distributed through legitimate applications. Techniques such as locality-sensitive hashing have been developed to efficiently identify shared components across large application corpora \cite{crussell2014andarwin}. Dedicated library detection systems \cite{libscout,libid,libscan,libd,libmd} identify recurring library code across versions and code transformations, as do graph-similarity approaches including SAD \cite{sadcdg}, MatchScope~\cite{matchscope} and LibHunter~\cite{libhunter24}; these systems match reused libraries, whereas our task assigns applications to proxy families. This challenge is particularly relevant to residential proxy attribution, as proxy families often embed common networking SDKs that can obscure family-specific code patterns. However, in the residential proxy context, this pervasive SDK reuse is not merely a complication, but the property that may make structural attribution tractable, as the same provider-specific SDK code reappears across many host applications.

\section{Problem Statement}
We formalize the task of attributing Android applications to residential proxy families and describe the challenges that motivate our approach.

\subsection{Task definition}
Let $\mathcal{A}$ denote the space of Android application packages (APKs) and let $\mathcal{F} = \{f_1, \ldots, f_K\}$ be a set of $K$ known residential proxy families. Given a labeled training corpus $\{(a_i, y_i)\}_{i=1}^{N}$ where $a_i \in \mathcal{A}$ and $y_i \in \mathcal{F}$, the goal is to learn a classifier that correctly attributes an unseen APK to its proxy family, i.e., for an unseen sample $(a, y)$ drawn from the same distribution, we seek a prediction $(h \circ \phi)(a) = y$. We formulate attribution as a single-label task: the 41 builds that carry SDKs of several providers are kept out of training (Section~\ref{sec:data}). Each family is a commercial provider, and successive SDK generations of one provider share its label.

We decompose this into two stages. First, a feature extraction function $\phi : \mathcal{A} \to \mathbb{R}^d$ maps each APK to a fixed-dimensional feature vector by disassembling the application, extracting control-flow and function-call graphs, and applying the Weisfeiler-Lehman subtree feature map to produce a $d$-dimensional vector encoding the structural patterns present in the application. Second, a classifier $h : \mathbb{R}^d \to \mathcal{Y}$ maps the feature vector to a label, where $\mathcal{Y} = \mathcal{F}$ for attribution and $\mathcal{Y} = \mathcal{F} \cup \{\mathtt{non\_proxy}\}$ when a $\mathtt{non\_proxy}$ class is trained. The composed function $h \circ \phi : \mathcal{A} \to \mathcal{Y}$ is the attribution pipeline. We evaluate performance using macro-averaged F1, which weights each family equally regardless of class size, under 5-fold DEX-grouped cross-validation that prevents any DEX file from appearing in both training and test partitions, and under a stricter grouping by host lineage (Section~\ref{sec:data_groups}).

This is a multi-class classification problem with $K = 8$ proxy families and a ninth \texttt{non\_proxy} class representing applications outside the known families; the feature dimension is $d = 2{,}048$ (1,024 WL features per graph type). We evaluate two regimes that differ in their input distribution. Under the attribution regime, inputs are APKs known to belong to one of the $K$ proxy families, and the classifier's predicted label is compared against the ground-truth family, with macro-averaged F1 as a measure of attribution performance. Under the detection regime, inputs may belong to one of the $K$ known proxy families or fall outside them, and we use the predicted label as a detection signal for the eight identified families; we evaluate it with the nine-class classifier on a sampled \texttt{non\_proxy} class and, separately, by rejecting independently collected applications. Because the \texttt{non\_proxy} class is a finite sample rather than a model of all unknown applications, the pipeline emits a label only when the maximum class probability $s(a) = \max_{c \in \mathcal{Y}} p_h(c \mid \phi(a))$ reaches a threshold $\tau$ and otherwise rejects the APK as unknown. A rejection is therefore distinct from a confident \texttt{non\_proxy} prediction, and since $s(a)$ is not calibrated, it measures the model's uncertainty rather than establishing whether an application contains proxy code. Our main rejection results use the eight-class model ($\mathcal{Y} = \mathcal{F}$); Section~\ref{sec:opensetproto} also reports the nine-class model.

\subsection{Challenges}
Several properties of the residential proxy ecosystem make this attribution problem non-trivial:

\begin{itemize}
    \item \textbf{Network-level opacity.} Backconnect traffic from proxy applications is typically routed through intermediate virtual private servers (VPSes), obscuring the identity of the proxy provider at the network layer. Network-level forensic artifacts alone are often insufficient for reliable family attribution.

    \item \textbf{SDK embedding and code reuse.} Residential proxy functionality is typically distributed as a monetization SDK embedded within otherwise-unrelated host applications. Multiple distinct APKs may embed the same SDK, and different families may share common third-party libraries for networking, analytics, or advertising. This extensive code reuse makes simple string or hash signatures unreliable, as they may match shared library code rather than family-specific SDK code.

    \item \textbf{Class imbalance.} Proxy families are unevenly represented in the corpus, which may reflect differences in SDK distribution and adoption. Some families account for thousands of application builds while others appear in only a few dozen, creating severe class imbalance that can bias classifiers toward larger families and motivates macro-averaged metrics.
\end{itemize}

\subsection{Approach overview}
Our approach addresses these challenges by extracting structural graph representations from application code, specifically control-flow graphs and function-call graphs, and using Weisfeiler-Lehman graph kernel features as input to machine learning classifiers. Rather than relying on surface-level string or byte-pattern matching, graph-based features capture both the code elements and their surrounding structure; how stable they are under repackaging and identifier renaming depends on the node labels, which we measure in Section~\ref{sec:robustness}. The classifier's predictions allow analysts to attribute an application to a proxy family and to select it for further reverse engineering.

This pipeline has direct applicability beyond the research corpus studied in this paper. Commercial app stores such as Google Play and the Amazon Appstore host large volumes of applications, a fraction of which may embed proxy SDKs without adequate disclosure to users. A validated attribution pipeline could be applied to app store corpora to identify and flag applications containing known proxy SDK families for review, supporting both platform enforcement and threat intelligence efforts; given the false-attribution rate observed on independently collected applications, we treat its predictions as triage signals rather than definitive evidence of proxy functionality (Section~\ref{sec:opensetproto}).

\section{Dataset}\label{sec:data}
We study the collection of residential proxy APKs assembled by Mi et al.\ \cite{mi2021your}, who identified 1,701 Android applications whose network traffic connected to backconnect infrastructure of four major residential proxy providers: IPNinja, Luminati (now BrightData), Monkeysocks, and Oxylabs. Their labeling was based on observed network connections to known provider infrastructure, providing ground truth family labels for each APK. Of the 1,701 hashes provided, 1,629 were available for download and form our original dataset while the other 72 were sourced solely from APKMonk and are not known to be accessible via Androzoo, VirusTotal, or other malware repositories. For this reason, we restrict our analyses to APK samples which are currently obtainable from those repositories with an API key.

\subsection{Dataset expansion via DEX reuse}
Residential proxy applications are typically distributed as monetization SDKs embedded within host applications. This means that the same SDK code -packaged as one or more DEX (Dalvik Executable) files -often appears across many distinct APKs. We leveraged this property to expand the dataset; using the VirusTotal platform, we identified DEX file hashes shared across APKs within each family, then used these shared DEX hashes as pivot points to discover additional applications containing the same SDK code. These newly identified applications were downloaded from ApkMonk and Androzoo \cite{allix2016androzoo}, expanding the dataset from 1,629 to 3,365 APKs.

DEX reuse enumeration is a reproducible technique for expanding labeled corpora in domains with heavy SDK embedding, such as the residential proxy ecosystem where a small set of SDK implementations is packaged into many host applications.

\paragraph{Limits of DEX-based label propagation} We unpacked every application in the corpus and scanned its DEX strings and native libraries for SDK markers of 27 proxy providers, established by reverse engineering the SDKs (namespaces, class names, protocol strings and library names). Of the 1,736 pivot-discovered applications, 675 (39\%) carry no marker of their assigned provider, and every one of them is linked to the labeled set only through a DEX file that holds a third-party library. Multidex builds place whole libraries in their own byte-identical DEX files, either as secondary \texttt{classes}$N$\texttt{.dex} files or under \texttt{assets/}: 292 of 303 unconfirmed Monkeysocks applications are linked through an advertising-mediation bundle that a single original application also ships, 100 of 177 unconfirmed Oxylabs applications through one Facebook Ads DEX file, and the 195 unconfirmed Luminati applications through library DEX files such as RxJava and Joda-Time. Labels propagated this way describe the library, not the provider. The original applications behave differently: their labels come from network observation, and markers are absent from only 0.1\% of original Monkeysocks applications, 21\% of Luminati and 93\% of Oxylabs, whose SDK does not keep a stable namespace.

\paragraph{Validated corpus} The released corpus applies four rules. (i) Original applications keep their network-derived label, and whether a marker confirms it is recorded. (ii) A DEX-pivot application is kept only if it carries its own provider's marker; the 675 others are dropped. The marker corroborates the propagated label but does not show that the pivot DEX file itself contained provider code, and markers are supporting evidence rather than independent ground truth: only the original applications are labeled by observed network behavior. (iii) Further builds of labeled packages, taken from the AndroZoo version history \cite{allix2016androzoo}, are added when their markers name exactly one provider. (iv) Four providers are added from a seed set built independently of the corpus: we enumerate Android applications reported as contacting the command-and-control domains published for IPIDEA, Popa and other providers in public threat reports, retrieve them and their other versions by package name from AndroZoo, APKPure and Aptoide, and label each build by the provider markers it contains. Builds with markers of two or more providers form a separate multi-label set of 41 applications, and providers with fewer than 15 single-provider builds (Pawns, Honeygain, Massive, LumiApps, Infatica and Repocket, 27 builds in total) are released but not used as classes. Bright Data and Luminati are one provider, and builds of the Netas engine are labeled Popa, whose SDK it became.

Before validation, the expanded dataset was class-imbalanced, as Luminati and Monkeysocks together comprised 91\% of samples, while IPNinja contributed only 35 samples. This imbalance is inherited from the Mi et al.\ corpus and preserved by our DEX-pivot expansion. Table~\ref{tab:dataset} summarizes the class distribution of the validated corpus: 3,781 applications in eight classes, 94\% of them confirmed by a marker of their assigned provider. Oxylabs remains the exception, at 19\%, and its applications are kept on the strength of their network-derived labels. In sum, the 1,629 original applications and 1,061 confirmed pivots give 2,690, and 717 version builds and 374 builds from command-and-control domains bring the corpus to 3,781. We release the corpus as SHA-256 hashes with this provenance (Section~\ref{sec:openscience}). The corpus remains imbalanced: Luminati and Monkeysocks hold 85\% of the applications, and Traffmonetizer, CashRaven and IPNinja fewer than 40 each.

\begin{table*}[htbp]
\centering
\small
\setlength{\tabcolsep}{4pt}
\begin{tabular}{lrrrrrrrr}
\toprule
& \multicolumn{5}{c}{\textbf{Applications by label source}} & \multicolumn{3}{c}{\textbf{Dependence and confirmation}} \\
\cmidrule(lr){2-6}\cmidrule(lr){7-9}
\textbf{Provider} & \textbf{Mi et al.} & \textbf{DEX pivot (drop.)} & \textbf{Versions} & \textbf{C2} & \textbf{Total} & \textbf{DEX gr.} & \textbf{Lineage gr.} & \textbf{Marker} \\
\midrule
Luminati / Bright Data & 757 & 715 (195) & 525 & 88 & 2,085 & 1,109 & 238 & 0.922 \\
Monkeysocks & 770 & 322 (303) & 27 & 6 & 1,125 & 33 & 27 & 0.999 \\
IPIDEA & & & 90 & 115 & 205 & 122 & 22 & 1.000 \\
Popa & & & 53 & 129 & 182 & 141 & 63 & 1.000 \\
Oxylabs & 80 & 11 (177) & & & 91 & 74 & 22 & 0.187 \\
CashRaven & & & 18 & 21 & 39 & 37 & 6 & 1.000 \\
IPNinja & 22 & 13 (0) & 2 & & 37 & 22 & 19 & 1.000 \\
Traffmonetizer & & & 2 & 15 & 17 & 15 & 9 & 1.000 \\
\midrule
\textbf{Total} & \textbf{1,629} & \textbf{1,061 (675)} & \textbf{717} & \textbf{374} & \textbf{3,781} & \textbf{1,552} & \textbf{404} & \textbf{0.937} \\
\bottomrule
\end{tabular}
\caption{Corpus by provider and label source. DEX pivot: applications found through a DEX file shared with a labeled application and kept because they carry their provider's marker; in parentheses, those dropped because they do not and are linked only through a library DEX file. Versions: other builds of labeled packages; C2 domains: builds of applications reported as contacting published command-and-control domains. DEX and lineage groups: connected components under the two groupings of Section~\ref{sec:data_groups} (over the 3,777 applications with extracted graphs; one DEX group and two lineage groups span providers, so per-provider counts are not additive). Marker: share of applications in which a marker of the assigned provider is found.}
\label{tab:dataset}
\end{table*}

Corpus totals count every labelled application we obtained. Feature extraction succeeds for all but four proxy applications and for 996 of the 1,000 $\mathtt{non\_proxy}$ applications, so the evaluated counts reported in Section~\ref{sec:results} are 1,629 original, 3,777 over the eight providers, and 4,773 with the $\mathtt{non\_proxy}$ class.

\paragraph{Applications containing multiple providers} Among the original and DEX-pivot applications, the marker scan finds no second provider. The seed set shows that this does not hold in general: 41 builds of applications reached through command-and-control domains, or of their other versions, carry the SDKs of two to four providers, for example Packet SDK (IPIDEA), Popa, Pawns and Traffmonetizer in a single football-fixtures application. Such applications are kept out of single-label training and released as a separate set.

\paragraph{Background sets} Two sets of applications without a known proxy SDK serve as negatives. The first, used as a trained \texttt{non\_proxy} class, holds 996 analyzable applications from an earlier collection of 500 benign and 500 malicious Google Play applications. The second, used only for open-set evaluation and never for training, is a uniform random sample of 4,888 Google Play applications from AndroZoo \cite{allix2016androzoo}. Scanning every one of them with the markers above finds a provider string in a single application, which we exclude, leaving 4,887. Marker recall is below one for some providers, so neither set is guaranteed to be proxy-free, and proxy predictions on them are counted as false positives, which overstates the false-positive rate by at most the prevalence of proxy SDKs in each set.

\subsection{Two leakage controls}\label{sec:data_groups}
All evaluations use five-fold grouped cross-validation with one of two groupings. \textbf{DEX grouping} joins applications that share any top-level DEX file, transitively, so that no DEX file appears in both training and test partitions; it is the protocol of prior work on this corpus. Recomputed from the restored APKs, it is considerably coarser than the grouping reported for the earlier version of the corpus: the 1,092 Monkeysocks applications common to both versions fall into 28 groups rather than 350, and the 1,472 common Luminati applications into 553 rather than 1,273. The earlier DEX-grouped figures therefore placed applications that share a DEX file on both sides of some splits, and this correction may contribute to the difference between the two versions under DEX grouping. DEX grouping does not separate applications of the same host lineage that share no DEX file, such as successive builds of one developer's application. \textbf{Lineage grouping} additionally joins applications that share a package name or whose identifier-seeded WL vectors of the function-call graph have a cosine similarity of at least 0.99, again transitively. It yields 404 groups for the 3,777 evaluated applications of the eight providers, against 1,552 DEX groups; the largest lineage group holds 862 applications, all of them Monkeysocks (Table~\ref{tab:dataset}). Lineage grouping is stricter than necessary for some providers and is not a temporal split; Section~\ref{sec:res_limits} reports one separately. We report both protocols throughout, because the difference between them indicates how strongly a result depends on shared code and closely related host applications.

\section{Methodology}\label{sec:methodology}
This section describes our pipeline for extracting structural features from Android residential proxy applications and the classification framework used for family attribution. Figure~\ref{fig:pipeline} provides an overview of the pipeline.

\begin{figure*}[htbp]
    \centering
    \resizebox{\linewidth}{!}{
\begin{tikzpicture}[
  font=\scriptsize,
  node distance=2.2mm and 3.4mm,
  box/.style={draw, rounded corners=1pt, align=center, minimum height=5.6mm,
              inner xsep=2pt, fill=blue!4},
  gbox/.style={box, fill=green!8},
  ybox/.style={box, fill=black!4},
  ar/.style={-{Latex[length=1.4mm]}, thin},
  dar/.style={ar, dashed}
]
\node[box] (apk) {APK};
\node[box, right=of apk] (dis) {Androguard\\\tiny disassembly};
\node[box, above right=0.4mm and 3.4mm of dis] (cfg) {CFG};
\node[box, below right=0.4mm and 3.4mm of dis] (fcg) {FCG};
\node[box, right=8mm of dis] (wl) {WL kernel\\\tiny $H{=}2$, $d{=}1{,}024$};
\node[box, right=of wl] (wlv) {WL vector\\\tiny $2{,}048$-d};
\node[box, right=of wlv] (cat) {Feature\\\tiny concat};
\node[box, above right=0.4mm and 3.4mm of cat] (clf) {Classifier\\\tiny SGD / RF / XGB};
\node[box, below right=0.4mm and 3.4mm of cat] (xai) {SHAP / LIME\\\tiny explainability};
\node[gbox, right=of clf] (lab) {Family\\label};
\node[gbox, right=of xai] (meth) {Influential\\methods};

\draw[ar] (apk) -- (dis);
\draw[ar] (dis) -- (cfg); \draw[ar] (dis) -- (fcg);
\draw[ar] (cfg) -| ($(wl.north)+(0,2mm)$) -- (wl.north);
\draw[ar] (fcg) -| ($(wl.south)-(0,2mm)$) -- (wl.south);
\draw[ar] (wl) -- (wlv);
\draw[ar] (wlv) -- (cat);
\draw[ar] (cat) -- (clf); \draw[ar] (cat) -- (xai);
\draw[ar] (clf) -- (lab); \draw[ar] (xai) -- (meth);
\end{tikzpicture}}
    \caption{Overview of the feature extraction and classification pipeline. Each APK is disassembled to extract CFG and FCG representations, which are converted to fixed-dimensional feature vectors via the WL kernel. The combined feature vectors are used for classification.}
    \label{fig:pipeline}
\end{figure*}

\subsection{Graph extraction}\label{sec:graph_extraction}
We used Androguard to disassemble each APK and extract two complementary graph representations from the Dalvik bytecode.

A control-flow graph (CFG) represents the intra-method branching structure of a single method; nodes correspond to basic blocks -maximal sequences of instructions with a single entry point and single exit point -and directed edges represent control transfers such as conditional branches, unconditional jumps, and exception handlers. Each method in the application produces one CFG, and we aggregate per-method CFGs into a single per-application CFG by taking their union. CFGs capture the fine-grained logic of individual routines, including loop structures, conditional dispatch, and error handling patterns.

A function-call graph (FCG) represents the inter-method call relationships across the entire application; nodes correspond to methods and directed edges represent call sites, where an edge $(m_1, m_2)$ indicates that method $m_1$ contains a call instruction targeting method $m_2$. FCGs capture the higher-level architectural organization of the application, including how SDK entry points, utility functions, and application logic are connected.

\begin{figure*}[htbp]
    \centering
    \includegraphics[width=0.75\linewidth]{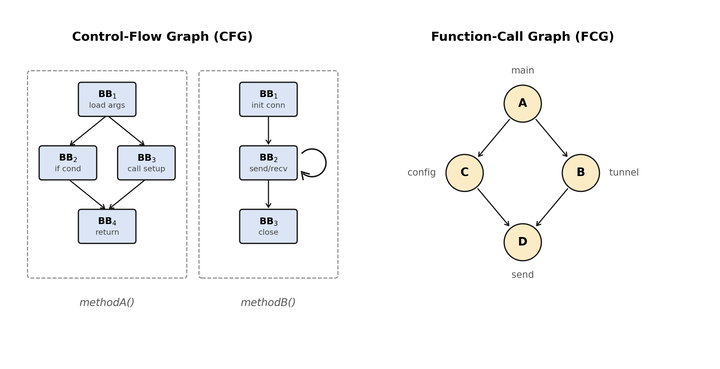}
    \caption{Illustration of the two graph representations extracted from each APK. Left: control-flow graphs capture intra-method branching structure, where nodes are basic blocks and edges are control transfers. Each method produces a separate CFG. Right: the function-call graph captures inter-method relationships, where nodes are methods and edges are call sites.}
    \label{fig:cfg_fcg}
\end{figure*}

These two representations are complementary; CFGs encode what happens within each method, while FCGs encode how methods relate to one another. Both are directed graphs, and we preserve edge direction throughout our analysis. Figure~\ref{fig:cfg_fcg} illustrates both representations for a simplified example.

\subsubsection{Data pre-processing}
To focus feature extraction on application-specific code rather than platform boilerplate, we performed one pre-processing step:
\begin{itemize}
    \item We filtered out graph nodes corresponding to standard library and common third-party packages to isolate the proxy SDK and host application code. The excluded package prefixes are:
    \begin{itemize}
        \item \texttt{java.*}, \texttt{javax.*}
        \item \texttt{android.*}, \texttt{androidx.*}
        \item \texttt{kotlin.*}
        \item \texttt{com.google.*}
        \item \texttt{org.apache.*}, \texttt{org.json.*}, \texttt{org.xml.*}
        \item \texttt{sun.*}, \texttt{dalvik.*}
    \end{itemize}
\end{itemize}

\subsection{WL feature vector construction}
To generate fixed-dimensional feature vectors from the variable-size graph representations, we applied the Weisfeiler-Lehman subtree kernel described in Section~\ref{sec:graph_kernels}. Our implementation differs from the standard WL formulation in two respects, both motivated by the directed nature of program graphs: (1) direction-aware neighbor aggregation that hashes predecessor and successor neighborhoods separately, and (2) a signed hash projection that maps WL labels into a fixed-dimensional vector. Algorithm~\ref{alg:wl} gives the full procedure.

\begin{algorithm}[H]
\caption{Direction-aware WL feature extraction}\label{alg:wl}
\begin{algorithmic}[1]
\Require Directed graph $G = (V, E)$, iterations $H$, dimension $d$
\Ensure Feature vector $\mathbf{x} \in \mathbb{R}^d$
\State $\mathbf{x} \gets \mathbf{0} \in \mathbb{R}^d$
\ForAll{$v \in V$}
    \State $\ell^{(0)}(v) \gets \text{\textsc{Hash}}(\text{label}(v))$
    \State $(i, s) \gets \text{\textsc{SignedHash}}(\ell^{(0)}(v), d)$
    \State $\mathbf{x}[i] \gets \mathbf{x}[i] + s$
\EndFor
\For{$h = 0$ \textbf{to} $H - 1$}
    \ForAll{$v \in V$}
        \State $\ell^{(h+1)}(v) \gets \text{\textsc{Hash}}\!\big(\ell^{(h)}(v),$
        \Statex \hspace{4em}$\{\!\!\{ \ell^{(h)}(u) : u \in \mathcal{N}^{-}(v) \}\!\!\},$
        \Statex \hspace{4em}$\{\!\!\{ \ell^{(h)}(w) : w \in \mathcal{N}^{+}(v) \}\!\!\}\big)$
        \State $(i, s) \gets \text{\textsc{SignedHash}}(\ell^{(h+1)}(v), d)$
        \State $\mathbf{x}[i] \gets \mathbf{x}[i] + s$
    \EndFor
\EndFor
\State \Return $\mathbf{x}$
\end{algorithmic}
\end{algorithm}

First, we use direction-aware neighbor aggregation. Rather than hashing the combined neighborhood of each node, we hash predecessor and successor neighborhoods separately. For a node $v$ at iteration $h$, the relabeling step becomes:
\begin{multline}
\ell^{(h+1)}(v) = \text{\textsc{Hash}}\!\Big(\ell^{(h)}(v),\; \{\!\!\{ \ell^{(h)}(u) : u \in \mathcal{N}^{-}(v) \}\!\!\},\\
\{\!\!\{ \ell^{(h)}(w) : w \in \mathcal{N}^{+}(v) \}\!\!\}\Big)
\end{multline}
where $\mathcal{N}^{-}(v)$ and $\mathcal{N}^{+}(v)$ denote the predecessor and successor neighborhoods of $v$, respectively. This preserves the distinction between callers and callees in FCGs and between branch sources and branch targets in CFGs, information that would be lost under undirected aggregation.

Second, we use a signed hash projection to map the (potentially unbounded) set of WL labels into a fixed-dimensional vector. Each relabeled node is hashed to determine both a bucket index $i \in \{0, \ldots, d-1\}$ and a sign $s \in \{+1, -1\}$; the feature vector is updated as $\mathbf{x}[i] \mathrel{+}= s$. This is a form of feature hashing  that reduces collision artifacts compared to unsigned projection, as colliding labels with opposite signs tend to cancel rather than accumulate \cite{weinberger2009feature}.

We apply $H = 2$ WL iterations with a hash dimension of $d = 1{,}024$ for each graph type, and concatenate the CFG and FCG vectors into a $2{,}048$-dimensional representation. The choice of 2 iterations captures structural patterns up to 2-hop neighborhoods, where iteration 0 encodes individual node identities (method names or basic block signatures), iteration 1 encodes each node's immediate neighborhood, and iteration 2 encodes 2-hop patterns. This is sufficient to distinguish most structurally distinct methods while keeping the feature space compact. The hash dimension of 1,024 was selected to balance expressiveness against sparsity; larger dimensions yielded diminishing returns in preliminary experiments while increasing computational cost.

We use two initial colorings. The \textsc{identifier} coloring uses the method signature, which makes the features strongly discriminative but changes completely under identifier renaming. The \textsc{combined} coloring replaces the signature with a renaming-invariant descriptor of the method: its signature with developer type and method names removed, the set of framework APIs it calls, and a hash of its opcode multiset.

\subsection{Baselines}\label{sec:method_baselines}
\paragraph{Drebin features} Drebin \cite{arp2014drebin} represents an application as a binary vector of manifest and code features. Only the bytecode of our corpus is retained, so we use the code sets: framework API calls (methods referenced but not defined by the application), the Drebin list of suspicious API calls, and network addresses found in DEX strings. Features present in at least five applications form the vocabulary (185,089 features for the eight-provider corpus).

\paragraph{MaMaDroid features} MaMaDroid \cite{mariconti2017mamadroid} abstracts every call to the family (11 states) or package (here the 120 most common API packages, plus self-defined and obfuscated) of the callee and represents the application by the row-normalized transition matrix of the resulting Markov chain, which we build from the call edges of the FCG.

\paragraph{Library detection} A library detector answers the attribution question directly if each provider's SDK can be used as a reference library. We use LibScan \cite{libscan} at library level. For every fold and family we disassemble training-fold applications in which the family's SDK namespace is present, reassemble the namespace classes into a reference DEX, and keep up to 15 distinct SDK versions; an application is attributed to the family of its best-matching reference and is rejected if nothing matches. References come only from training folds, the condition that corresponds to the other baselines, and they cannot be built for families whose namespace is rarely recoverable, which is itself a property of library detection on this corpus.

\subsection{Classification}
We evaluate two classifiers chosen to span the primary model families used for tabular classification: a linear model and a bagged-tree ensemble. This selection covers models with qualitatively different inductive biases, so that the reported performance reflects the feature representation rather than a single model family's fit to these features.
\begin{itemize}
    \item Logistic regression trained by Stochastic Gradient Descent (SGD), representing a linear model that operates directly in the feature space, with the $\ell_2$ strength chosen by an inner three-fold grouped cross-validation; for Drebin we also fit a linear SVM, as in the original work. Because per-feature standardization changes the linear model's accuracy substantially for some representations, we report the better of raw and standardized features.
    \item Random Forest (RF) with 200 to 500 estimators, an ensemble of decision trees that captures non-linear interactions between features.
\end{itemize}

All classifiers are evaluated using 5-fold DEX-grouped cross-validation. We compute connected components over SHA-256 hashes of all DEX files in the dataset, assigning each APK to a connected component such that any two APKs sharing a DEX file belong to the same group; we then use GroupKFold so that no DEX file appears in both training and test partitions of any fold. This prevents the classifier from achieving high test scores by memorizing specific DEX artifacts shared across APKs and forces it to learn family-level structural patterns instead. The heavy SDK reuse documented in Section~\ref{sec:graph_extraction} makes such memorization a plausible failure mode under standard stratified splits. As a second leakage control, we repeat every evaluation under lineage grouping (Section~\ref{sec:data_groups}). We report the mean of each metric across the folds of three repetitions of the five-fold split (one repetition for Drebin and for the experiments of Section~\ref{sec:res_limits}). Due to class imbalance, we upsample minority classes within each training fold to match the majority class size. We report macro-averaged F1 score as the primary metric, as it weights each class equally regardless of size and is therefore sensitive to performance on minority classes such as IPNinja. For thresholded classification, a sample is assigned one of the trained labels only when the highest class confidence exceeds the selected threshold, and is otherwise rejected as unknown.

\subsection{Limits}
\paragraph{Unseen providers} To ask whether the features describe proxyware or only the providers in training, we train a binary proxy-against-background detector on all providers but one and the background class and measure, at a threshold giving a 1\% false-positive rate on held-out background applications, the fraction of the held-out provider's applications it flags.

\paragraph{SDK generations} Luminati rewrote its SDK during the corpus period. Generation~A (\texttt{io.topvpn}, \texttt{org.hola}) is a Java wrapper around a native proxy engine, delivered as a downloaded executable or an embedded runtime; generation~B (\texttt{io.lum.sdk}) ports the engine to Java with the same protocol field names, and generations C (\texttt{lumnet.sdk}) and D (\texttt{brdat.sdk}) rename and obfuscate B. We tag each Luminati application with the newest generation whose namespace it contains (1,721 A, 200 B to D; 164 untagged applications are left out) and, within DEX-grouped folds, remove the Luminati training applications of one side: A$\rightarrow$BCD trains on generation A and scores all B to D applications, BCD$\rightarrow$A the reverse. Because B to D are eight times smaller than A, we also train on a random subsample of A of the same size as the B to D training set in each fold. Besides recall on the held-out generation we report the share of the other providers' test applications predicted as Luminati, since high recall means little if the model calls everything Luminati.

\paragraph{Identifier renaming} We rename every developer-defined type and method in each application, consistently within the application, keeping framework identifiers, relabel the extracted graphs accordingly, re-vectorize them and score the \textsc{identifier} model trained on unmodified applications on the renamed ones, reporting the share of predictions that change.

\paragraph{Open set} A model trained on the providers is applied to the 4,887 marker-screened Play applications. We compare three novelty scores: the maximum class probability, the cosine similarity to the nearest class mean, and the largest decision value of a per-family one-class SVM. For each we report the AUROC of known against unknown applications and, at the threshold that retains 95\% of known applications, the false-positive rate on the Play sample and the precision implied at prevalence $\pi$, $\mathrm{PPV}(\pi)=\mathrm{TPR}\,\pi/(\mathrm{TPR}\,\pi+\mathrm{FPR}\,(1-\pi))$.

\section{Results}\label{sec:results}
We compared two classifiers -a linear model and Random Forest (RF) -on WL features (2,048 dimensions) and on the Drebin and MaMaDroid baseline representations, across both the eight providers of the validated corpus and its four original providers. Due to class imbalance, we upsampled minority classes within each training fold. All results use 5-fold DEX-grouped cross-validation, in which all APKs sharing a DEX file are constrained to the same fold so that no DEX file appears in both training and test partitions, are repeated under lineage grouping, and report macro-averaged F1 score.

\subsection{Classification Results}\label{sec:res_baselines}
Table~\ref{tab:baselines} reports macro-F1 for every representation under both leakage controls, over the eight providers and, for comparison with the earlier version of the corpus, over its four original providers with the validated labels.

\begin{table*}[htbp]
\centering
\small
\begin{tabular}{lcccccccc}
\toprule
& \multicolumn{4}{c}{\textbf{Eight providers}} & \multicolumn{4}{c}{\textbf{Four original providers}} \\
\cmidrule(lr){2-5}\cmidrule(lr){6-9}
& \multicolumn{2}{c}{\textbf{DEX grouping}} & \multicolumn{2}{c}{\textbf{Lineage grouping}} & \multicolumn{2}{c}{\textbf{DEX grouping}} & \multicolumn{2}{c}{\textbf{Lineage grouping}} \\
\cmidrule(lr){2-3}\cmidrule(lr){4-5}\cmidrule(lr){6-7}\cmidrule(lr){8-9}
\textbf{Representation} & \textbf{Lin.} & \textbf{RF} & \textbf{Lin.} & \textbf{RF} & \textbf{Lin.} & \textbf{RF} & \textbf{Lin.} & \textbf{RF} \\
\midrule
WL \textsc{identifier} & \textbf{0.950} & 0.826 & 0.722 & 0.433 & 0.977 & 0.886 & 0.896 & 0.592 \\
WL \textsc{combined} & 0.908 & 0.816 & 0.655 & 0.397 & 0.948 & 0.800 & 0.763 & 0.491 \\
MaMaDroid, family & 0.757 & 0.904 & 0.367 & 0.584 & 0.810 & 0.902 & 0.425 & 0.684 \\
MaMaDroid, package & 0.906 & 0.900 & 0.589 & 0.595 & 0.878 & 0.891 & 0.627 & 0.679 \\
Drebin (code) & 0.938 & \textbf{0.951} & \textbf{0.786} & \textbf{0.741} & \textbf{0.992} & \textbf{0.995} & \textbf{0.940} & \textbf{0.939} \\
\midrule
LibScan & \multicolumn{2}{c}{0.674 (WL 0.957)} & & & \multicolumn{2}{c}{0.651 (WL 0.962)} & & \\
\bottomrule
\end{tabular}
\caption{Macro-F1 over eight providers ($n=3{,}777$ applications with extracted graphs; Drebin $n=3{,}771$) and over the four original providers ($n=3{,}338$; Drebin $3{,}332$). Lin.: best of the tuned logistic regression on raw, $\ell_2$-normalized and per-feature standardized features (standardization statistics from the upsampled training fold); for Drebin, the better of logistic regression and a linear SVM. These choices are made on the reported folds, an optimistic selection applied to every representation alike; with the regularization chosen by inner cross-validation alone, WL \textsc{identifier} reaches 0.948 and 0.701. RF: random forest. WL rows use CFG and FCG features. LibScan: macro-F1 on held-out applications of one DEX-grouped repetition, with the WL \textsc{identifier} model on the same applications in parentheses. WL and MaMaDroid: three repetitions of five folds; Drebin: one.}
\label{tab:baselines}
\end{table*}

\paragraph{Attribution does not require graph features} Under DEX grouping, the forest on code-only Drebin features reaches 0.951 macro-F1 over the eight providers and the linear model 0.938, against 0.950 for the identifier-seeded WL model; MaMaDroid package features reach 0.906. On the four original providers Drebin reaches 0.992 and 0.995. No claim of superiority for graph-kernel features can rest on these numbers; what distinguishes the representations is how they degrade.

\paragraph{Validation changes graph features more than Drebin} On the four original providers, the earlier version of the corpus gave 0.998 (linear) and 0.996 (forest) for Drebin, 0.993 for the WL \textsc{identifier} model and 0.972 for MaMaDroid package features with a standardized linear model. With the validated labels, the stricter DEX grouping and the added version builds, Drebin is unchanged, WL \textsc{identifier} loses 2 points and MaMaDroid package 9. A plausible reason for MaMaDroid's loss is the 675 applications whose label came only through a shared library DEX file: a representation built from abstracted API calls can learn libraries, and a library was all those applications had in common with their assigned provider. The comparison also changes the grouping and adds version builds, so the contribution of label validation alone is not isolated. Marker-corroborated labels do not inflate the scores either. Scoring the cross-validated predictions separately by label source, applications labeled by observed network traffic reach 0.988 (WL \textsc{identifier}) and 0.994 (Drebin) macro-F1 under DEX grouping and 0.929 and 0.944 under lineage grouping, against 0.955 and 0.949, and 0.694 and 0.766, for marker-labeled applications; the network-labeled applications span only the four original providers, so part of their advantage is the easier task. Trained and tested on the network-labeled originals alone, the two models reach 0.984 and 0.988 under DEX grouping and 0.929 and 0.952 under lineage grouping, and with shuffled training labels every model falls to chance (0.09 to 0.11).

\paragraph{The new providers make attribution harder} Adding IPIDEA, Popa, CashRaven and Traffmonetizer lowers the best score of the WL and Drebin representations by 3 to 4 points under DEX grouping (Drebin 0.995 to 0.951, WL \textsc{identifier} 0.977 to 0.950), while MaMaDroid's is unchanged or slightly higher. Two of the four new classes, CashRaven and Traffmonetizer, have fewer than 40 applications, so each fold holds a handful of their test applications and a single error moves macro-F1 noticeably.

\paragraph{Host lineage carries much of the headline} Grouping by host lineage lowers every representation over the eight providers by 15 to 42 points. Drebin degrades least and is the most accurate representation under this grouping (0.786 linear, 0.741 forest), ahead of WL \textsc{identifier} (0.722). The forests lose most: the forest on WL \textsc{identifier} falls from 0.826 to 0.433 and on MaMaDroid family features from 0.904 to 0.584, which suggests that their DEX-grouped accuracy depended on recognizing near-copies of training applications. Lineage-grouped estimates also vary strongly between folds (the tuned WL \textsc{identifier} model has a fold standard deviation of 0.15), because the new providers fall into few lineage groups (six for CashRaven, nine for Traffmonetizer) and a fold can hold one lineage of a provider or none. On the four original providers, lineage grouping costs Drebin 5 points and WL \textsc{identifier} 8, as on the earlier corpus. We regard the lineage-grouped columns as the more realistic estimate for applications from developers the training set has not seen.

\paragraph{The WL colorings trade accuracy for robustness} The \textsc{combined} coloring, which is invariant to identifier renaming, is 4 points below \textsc{identifier} under DEX grouping and 7 under lineage grouping. Two further renaming-invariant colorings, a fuzzy signature and a structure-only coloring, are lower still under lineage grouping (0.597 and 0.574). \textsc{combined} is nonetheless the representation that transfers best across the Luminati rewrite (Section~\ref{sec:res_limits}).

\paragraph{What the graph adds} WL refinement yields no significant improvement over the initial colors of the two colorings we recommend. With FCG features over the eight providers (tuned linear model, two repetitions of five folds), the \textsc{identifier} model reaches 0.945 macro-F1 at $h=0$ and 0.941 at $h=2$ under DEX grouping, and 0.728 and 0.711 under lineage grouping; the \textsc{combined} model reaches 0.923 and 0.907, and 0.625 and 0.567. None of these differences is significant (corrected resampled $t$-test, $p>0.3$), and all point slightly against refinement. Refinement helps only when the initial color is weak: under DEX grouping it raises a framework-API coloring from 0.572 to 0.733 and a degree-only coloring from 0.488 to 0.664 ($p\le0.003$), which still leaves both well below the identifier-bearing colorings. For identifier-bearing representations, most of the discrimination therefore appears to come from which methods and framework calls an application contains rather than from how they are connected, consistent with Drebin's code-only features doing as well or better; a non-significant difference does not, however, establish that refinement contributes nothing.

\begin{figure}[t]
\centering
\includegraphics[width=\linewidth]{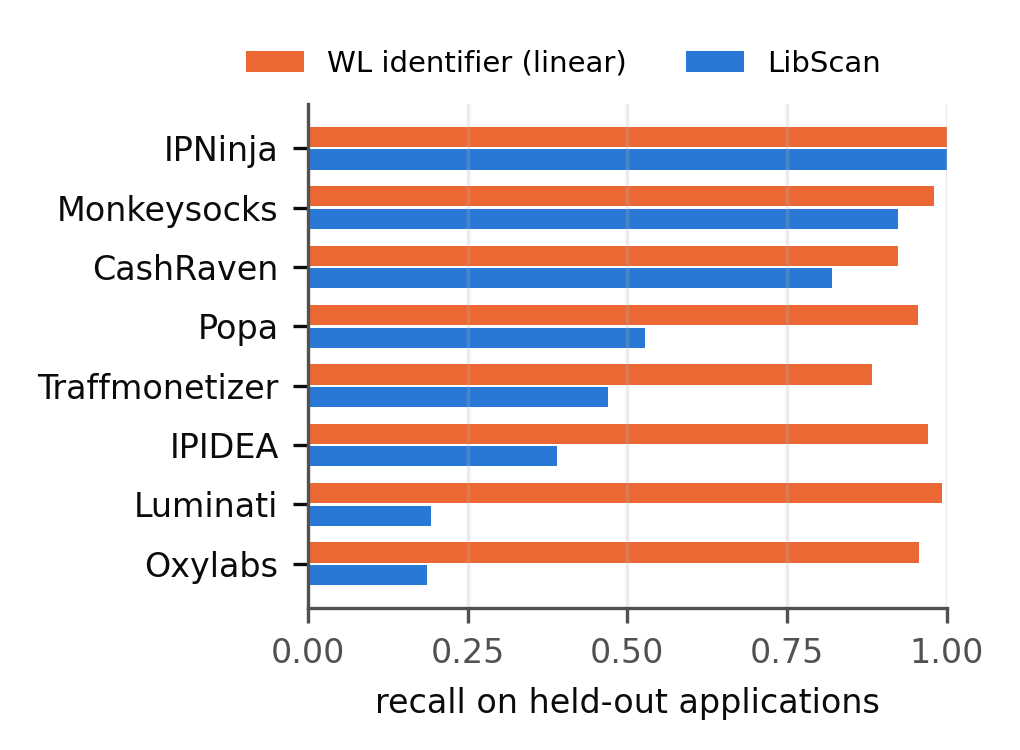}
\caption{Recall per provider of LibScan and of the WL \textsc{identifier} linear model on the same 1,167 held-out applications (one DEX-grouped repetition). Every LibScan error is a rejection.}
\label{fig:libscan}
\end{figure}

\paragraph{Library detection} Over the eight providers, LibScan reaches 0.674 macro-F1 on 1,167 held-out applications (up to 60 per provider per DEX-grouped fold), against 0.957 for the WL \textsc{identifier} model on the same applications; on the four original providers the figures are 0.651 and 0.962. On this sample LibScan never attributes an application to the wrong provider: every error is a rejection, and it detects a reference in 52\% of the evaluated applications. Recall follows whether the training folds yield usable references (Figure~\ref{fig:libscan}). It is high where an SDK keeps its namespace and few builds cover it, and low for Luminati, whose namespace changed between generations, for Oxylabs, whose namespace survives in under a fifth of applications, and for IPIDEA, for which the training folds yield only three references, since many IPIDEA builds carry the SDK as a native Go library rather than as bytecode. The WL model recalls 0.88 to 1.00 of every provider.

 A library detector is thus a high-precision, low-coverage complement to a learned model on this corpus, not a replacement for it.

\subsection{Open-set rejection protocol}\label{sec:opensetproto}
\begin{figure}[t]
\centering
\includegraphics[width=\linewidth]{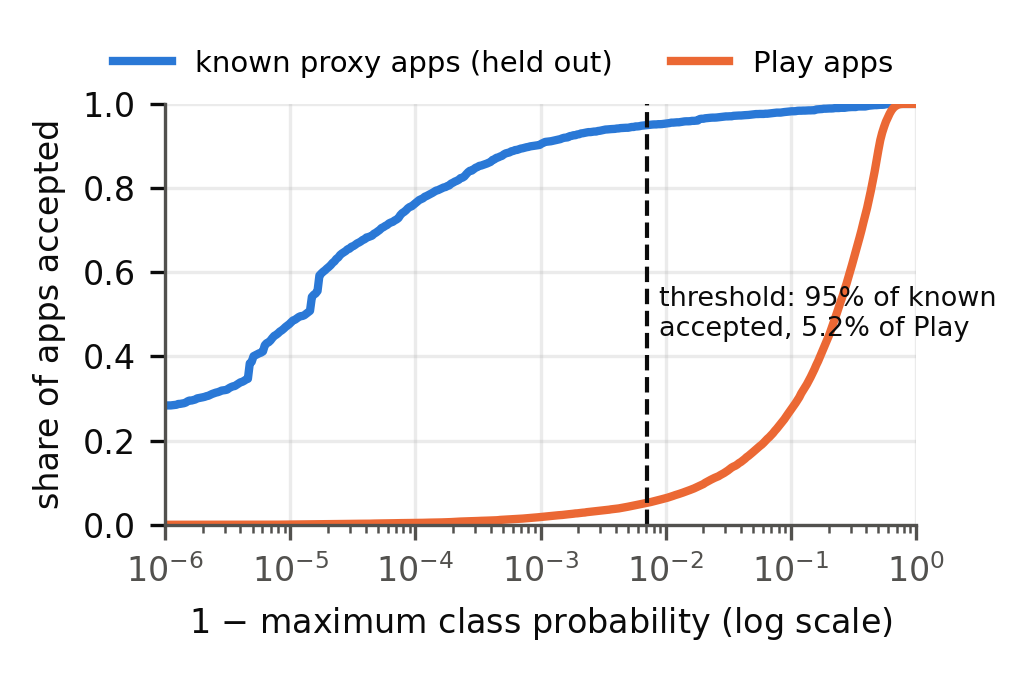}
\caption{Share of applications accepted as a function of the rejection threshold, for held-out known proxy applications and the 4,887 marker-screened Play applications (eight-class WL \textsc{identifier} linear model, DEX grouping; Play applications scored by every fold model). The dashed line is the threshold that retains 95\% of known applications.}
\label{fig:msp}
\end{figure}

\begin{figure}[t]
\centering
\includegraphics[width=\linewidth]{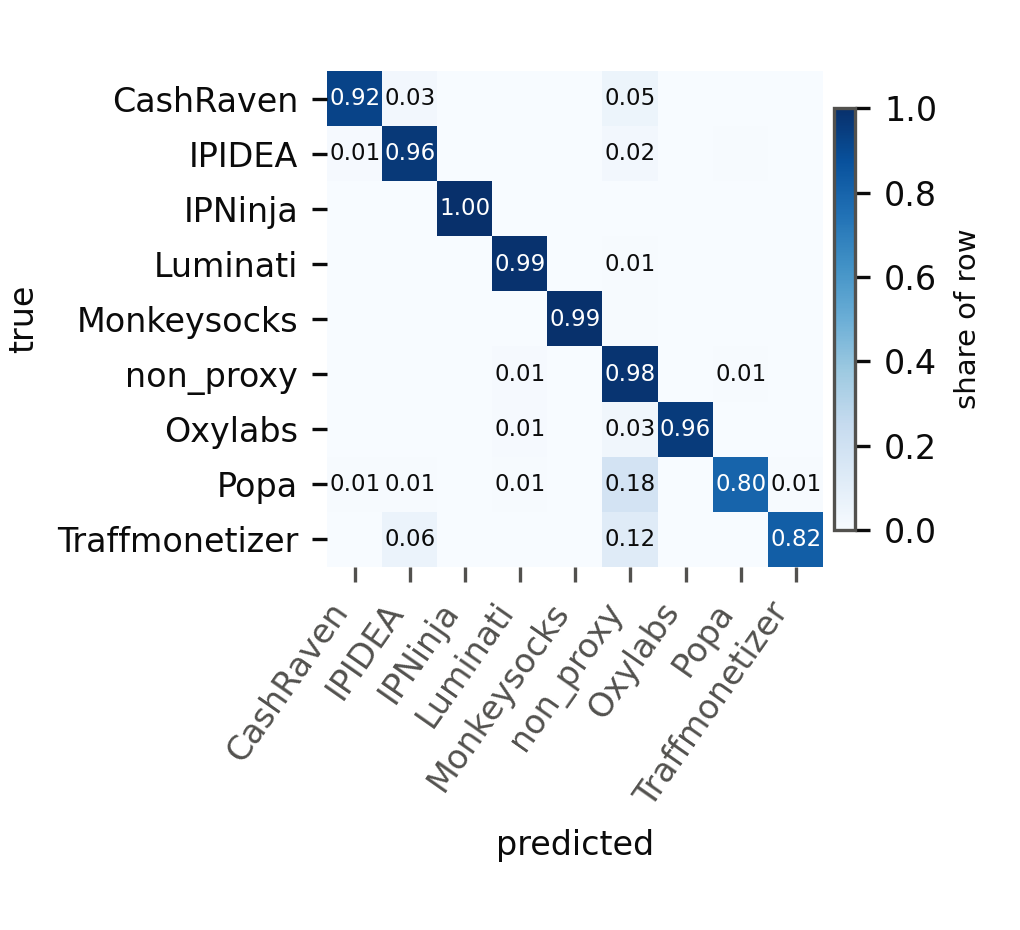}
\caption{Row-normalized confusion matrix of the nine-class WL \textsc{identifier} linear model (out-of-fold predictions, one repetition of DEX-grouped five-fold cross-validation, $n=4{,}773$). Empty cells are below 0.005.}
\label{fig:confusion}
\end{figure}

Classification with a trained $\mathtt{non\_proxy}$ class is closed-set: every test application belongs to a trained label, and the $\mathtt{non\_proxy}$ class is a finite sample of benign software rather than a model of all unknown applications. We therefore evaluate rejection separately, against applications drawn independently of the corpus, and specify the protocol here because the result depends on it. \\
Table~\ref{tab:openset} compares three novelty scores against the 4,887 marker-screened Play applications. Over the eight providers the maximum class probability is best, with an AUROC of 0.982 (Figure~\ref{fig:msp}); at the threshold retaining 95\% of known applications, 5.2\% of Play applications are still assigned a provider, which at a prevalence of one in a thousand implies a precision of 1.8\%. The best score is not stable across label sets: over the four original providers the maximum class probability drops to 0.889 and the one-class SVM is best, with a false-positive rate of 17\%, while over eight providers the one-class SVM is weakest. Training with the background applications as a ninth class does not remove the problem. The linear model then reaches 0.941 macro-F1 over nine classes, its largest errors assigning Popa and Traffmonetizer applications to $\mathtt{non\_proxy}$ (Figure~\ref{fig:confusion}), and assigns 18.5\% of the Play applications to a provider; abstaining below the confidence that keeps 95\% of proxy applications separates proxy from Play applications with an AUROC of 0.980 and lowers the Play false-positive rate to 7.3\%. A forest assigns only 2.8\% of Play applications to a provider, but at 0.852 macro-F1. Classifier output is therefore a triage signal for Play-scale scanning, not an attribution.

\begin{table}[htbp]
\centering
\small
\setlength{\tabcolsep}{2.5pt}
\begin{tabular}{lcccccc}
\toprule
& \multicolumn{3}{c}{\textbf{Eight providers}} & \multicolumn{3}{c}{\textbf{Four providers}} \\
\cmidrule(lr){2-4}\cmidrule(lr){5-7}
\textbf{Score} & \textbf{AUROC} & \textbf{FPR@95} & \textbf{PPV} & \textbf{AUROC} & \textbf{FPR@95} & \textbf{PPV} \\
\midrule
Max.\ prob. & \textbf{0.982} & \textbf{0.052} & \textbf{1.8\%} & 0.889 & 0.474 & 0.2\% \\
Class mean & 0.919 & 0.429 & 0.2\% & 0.965 & 0.245 & 0.4\% \\
OC-SVM & 0.922 & 0.593 & 0.2\% & \textbf{0.970} & \textbf{0.174} & \textbf{0.5\%} \\
\bottomrule
\end{tabular}
\caption{Known proxy applications against 4,887 marker-screened Play applications (WL \textsc{identifier}, linear, DEX grouping). Scores: maximum class probability, cosine similarity to the nearest class mean, per-provider one-class SVM. FPR@95: share of Play applications above the threshold that retains 95\% of known applications; PPV: precision at a prevalence of $10^{-3}$, assuming the measured rates transfer to that population.}
\label{tab:openset}
\end{table}

\subsection{Robustness and representation analysis}
\begin{figure}[t]
\centering
\includegraphics[width=\linewidth]{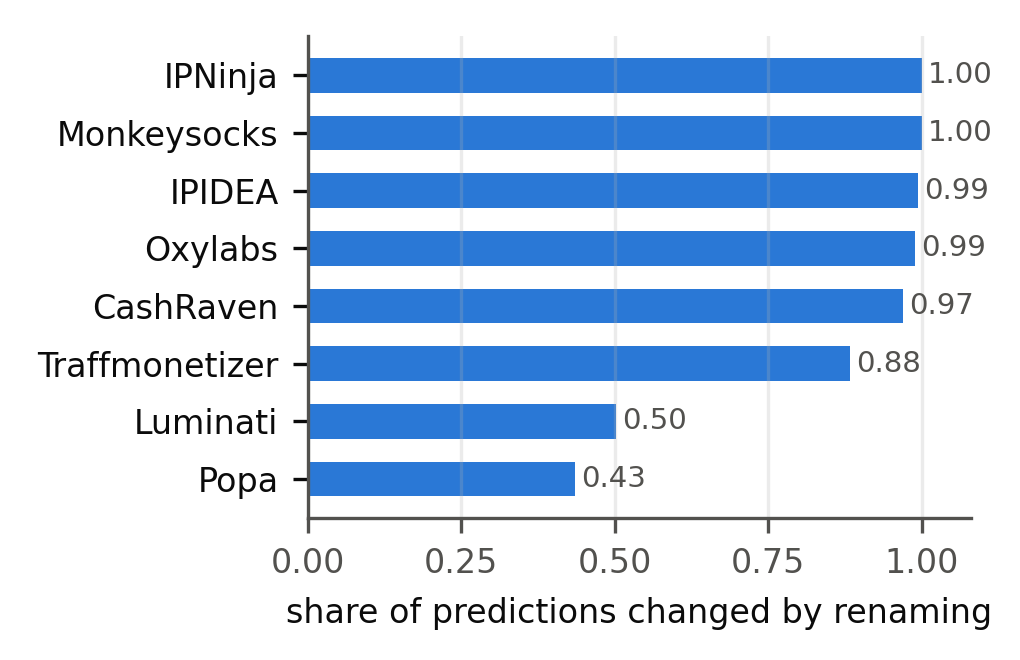}
\caption{Share of predictions of the WL \textsc{identifier} linear model (trained on unmodified applications, DEX grouping) that change when every developer-defined identifier is renamed.}
\label{fig:renaming}
\end{figure}

Renaming developer identifiers per application changes 70\% of the predictions of a WL \textsc{identifier} model trained on unmodified applications, which classifies 99\% of the unmodified versions correctly. The change is near-total for six providers and partial for Luminati and Popa (Figure~\ref{fig:renaming}).

 Over the four original providers the overall rate is lower, 37\%, but only because Luminati, 62\% of the applications there, almost never changes (0.2\%) while the other three providers change in 95 to 99\% of cases, consistent with renamed applications falling into the largest class. The \textsc{combined} coloring is unaffected by construction, at the cost in accuracy reported above.

\subsection{Limits}\label{sec:res_limits}

\paragraph{SDK generations} Table~\ref{tab:generations} reports how a model trained on one Luminati SDK generation attributes the other. Trained on the native-engine generation~A with as many applications as the Java generations offer, every representation still attributes 65 to 86\% of the Java builds to Luminati, and with all of generation~A 74 to 94\%. In the reverse direction, trained on the Java generations B to D, no representation attributes more than 43\% of native-engine builds. The SDK suggests why: the Java port keeps the wrapper's interface and protocol field names, which generation~A's bytecode already contains, whereas generation~A's proxy logic lives in native code that bytecode features do not see. Identifier features transfer worst in both directions (65\% and 6\%), because the rewrite changed the namespace, while the renaming-invariant \textsc{combined} coloring transfers best from the Java builds (43\%) and, trained on all of generation~A, to them (94\%). Forests on WL features reach full recall from all of generation~A, but only by predicting Luminati for 11 to 16\% of the other providers' test applications; the linear models stay below 4\%. A model trained on a provider's current SDK can therefore not be assumed to recognize its older, natively implemented builds.

\begin{table}[htbp]
\centering
\small
\setlength{\tabcolsep}{3pt}
\begin{tabular}{lccc}
\toprule
\textbf{Representation} & \textbf{A$\rightarrow$BCD} & \textbf{A (m.)$\rightarrow$BCD} & \textbf{BCD$\rightarrow$A} \\
\midrule
WL \textsc{identifier} & 0.739 & 0.648 & 0.063 \\
WL \textsc{combined} & 0.942 & 0.836 & \textbf{0.428} \\
MaMaDroid, package & 0.838 & 0.778 & 0.323 \\
Drebin (code) & 0.913 & \textbf{0.857} & 0.387 \\
\bottomrule
\end{tabular}
\caption{Recall on the held-out Luminati generation (linear models, DEX grouping, one repetition of five folds). A: native engine (1,721 applications); B to D: Java engine (200). A (m.): generation~A training applications subsampled to the size of the B to D training set. Other providers' test applications predicted Luminati: at most 3.7\%.}
\label{tab:generations}
\end{table}

\begin{figure}[t]
\centering
\includegraphics[width=\linewidth]{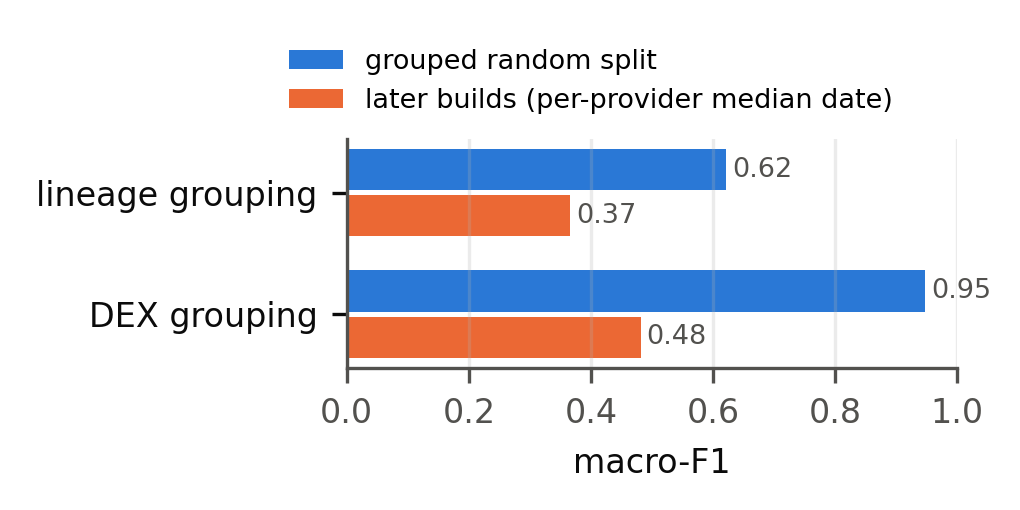}
\caption{Macro-F1 of the WL \textsc{identifier} linear model on the later half of each provider's builds (split at each provider's median VirusTotal first-submission date, test applications sharing a group with training removed) against grouped random splits of the same dated applications; seven providers.}
\label{fig:temporal}
\end{figure}

\paragraph{Within-provider temporal holdout} We dated applications by their first submission to VirusTotal, which covers 3,764 of the 3,777 applications; AndroZoo's first-scan date covers 2,670 and agrees with it to within five weeks for 84\% of the applications dated by both, and build timestamps are too often a placeholder to use. Providers occupy different periods (median first submission 2018.9 for Monkeysocks, 2019.1 for Luminati, 2022.8 for Popa and 2024.1 for IPIDEA), and the date alone predicts the provider with a macro-F1 of 0.24 against 0.12 for the class prior under lineage-grouped folds, so any split by date also changes the provider mix. A split at the corpus median date is therefore not informative here: four providers appear only after it. Instead, we split every provider at its own median date, trained on all earlier halves and tested on all later halves, removing test applications that share a group with a training application, and compared against grouped random splits of the same dated applications (seven providers; Traffmonetizer has too few applications). Under lineage grouping the WL \textsc{identifier} linear model reaches 0.366 macro-F1 on later builds against 0.622 on random splits, a gap of 26 points; under DEX grouping the gap is 47 points (0.481 against 0.948; Figure~\ref{fig:temporal}). Removing test applications that share a lineage with training leaves 459 later builds, mostly of host lineages first seen after the split, so the lineage-grouped gap measures attribution of new host lineages that also postdate training. The code does change within a provider: a linear model separates its later from its earlier builds with an AUROC of 0.85 (Luminati) to 0.87 (IPIDEA) under lineage grouping. Absolute values are lower than in Table~\ref{tab:baselines} because each model trains on half of the dated applications, and the dates are first-submission dates, not release dates.

\paragraph{Unseen providers} Table~\ref{tab:lofo} reports how often a binary proxy detector trained without one provider flags it at a 1\% false-positive rate on background applications. The two largest providers are almost never found: Monkeysocks in 0.5\% of cases and Luminati in 7\%, although their AUROC against background applications stays above 0.82. The smaller providers are found more often, in 9 to 53\% of cases, but most of their applications still go undetected. The features describe the providers in training rather than proxyware in general, and a deployment should expect to miss a provider absent from training.

\begin{table}[htbp]
\centering
\small
\begin{tabular}{lrcc}
\toprule
\textbf{Held-out provider} & \textbf{$n$} & \textbf{TPR at 1\% FPR} & \textbf{AUROC} \\
\midrule
Luminati & 2,085 & 0.073 & 0.829 \\
Monkeysocks & 1,125 & 0.005 & 0.907 \\
IPIDEA & 205 & 0.415 & 0.883 \\
Popa & 178 & 0.090 & 0.891 \\
Oxylabs & 91 & 0.110 & 0.810 \\
CashRaven & 39 & 0.385 & 0.889 \\
IPNinja & 37 & 0.216 & 0.887 \\
Traffmonetizer & 17 & 0.529 & 0.968 \\
\bottomrule
\end{tabular}
\caption{Share of the held-out provider's applications flagged by a proxy detector trained on the other seven, at a 1\% false-positive rate on held-out background applications, and AUROC against them (WL \textsc{identifier}, CFG and FCG, linear, DEX grouping; $n$ counts applications with extracted graphs).}
\label{tab:lofo}
\end{table}

\section{Discussion}\label{sec:discussion}
\subsection{DEX reuse}
We found that DEX reuse varies considerably across the four proxy families, which directly affects class imbalance in the expanded dataset as families with a higher degree of reuse see more applications added to their class. Table~\ref{tab:dataset} reports the resulting dependence structure. The 1,125 Monkeysocks applications fall into 33 DEX groups and 27 lineage groups, so the provider is in effect a handful of host lineages repeated many times; the 2,085 Luminati applications form 1,109 DEX groups but only 238 lineage groups, because successive builds of one application share no DEX file but remain one lineage; and IPNinja's 37 applications form 22 DEX and 19 lineage groups. These results lend insights into the code reuse environment and how monetization incentives impact the development environment for each family.

Residential proxy applications are distributed as monetization SDKs embedded within host applications, which is why DEX reuse rates are high. Family attribution reduces to detecting the presence of reused SDK code under varying packaging contexts; classifier performance of 0.950 macro-F1 under DEX-grouped evaluation (Table~\ref{tab:baselines}) indicates that SDK structural fingerprints persist across host applications, although the drop to 0.722 under lineage grouping shows that a substantial part of the signal depends on near-copies within a host lineage.

To verify that classification performance is not an artifact of DEX leakage between training and test folds, we conducted a DEX-grouped cross-validation experiment. Using union-find over SHA-256 hashes of all DEX files in the expanded dataset, we assigned each APK to a connected component such that any two APKs sharing a DEX file belong to the same group. We then replaced the standard 5-fold stratified cross-validation with GroupKFold, guaranteeing that no DEX file appears in both training and test partitions of any fold. The DEX-grouped results in Table~\ref{tab:baselines} therefore require generalization beyond exact DEX identity. Grouped evaluation does not, however, eliminate other correlations among related applications, which is why we also report lineage grouping.

\subsection{Ecosystem persistence and developer concentration}\label{sec:ecosystem}
Beyond the closed-world classifier evaluation, we apply the trained model to the current (APKPure) versions of each package in the validated corpus. The model is the linear WL \textsc{identifier} classifier over the eight providers and the $\mathtt{non\_proxy}$ class, and to limit false positives we accept a provider label only when the maximum class probability exceeds a threshold calibrated to a 1\% false-positive rate on the 4,887 marker-screened Play applications, which retains 91\% of known proxy applications. Every downloaded build is also scanned for the providers' SDK markers, and its SHA-256 recorded, so that each prediction can be checked against the build it was made on. Of the 1,508 corpus packages, 1,210 (80\%) are no longer listed on APKPure, including 975 of 1,019 Monkeysocks packages, most of them template applications. Of the 212 packages whose current build our pipeline analyzes, 139 (65.6\%) still classify as their original proxy family. This persistence varies by provider; Luminati retains 95 of 136 current packages, Popa 16 of 24 and Monkeysocks 15 of 31. The markers support the classifier on these packages: 138 of the 212 builds (65\%) carry their provider's SDK, and 129 of the 139 accepted predictions (93\%) are confirmed by it. Most current builds are byte-identical to a corpus build (126 of the 139 marker-bearing ones), so persistence here means largely that the applications have not been updated; with each package's publisher held out of training, the model still recognizes 76\% of the marker-bearing builds at 98\% precision, but only 1 of the 13 builds that changed since collection.

\sloppy A publisher-level pattern emerges when we pivot from packages to developer accounts. For each analyzed package we extract the publisher identity from the APKPure developer page, yielding 109 developers for 209 packages. We identify 23 developer accounts with at least two corpus packages whose current versions retain the original provider label, with no conflicting accepted prediction (Table~\ref{tab:retained_devs}), and enumerate each developer's full APKPure catalog to find applications not in our corpus. Of the resulting 771 candidates, 554 could be retrieved and analyzed, and the model assigns 197 of them to the developer's provider. The implication for threat intelligence is that flagging individual developer accounts, rather than individual APKs, can provide higher leverage, but verification shows how much the classifier's catalog predictions depend on which developer they come from.

Of the 197 flagged catalog applications, only 26 carry their provider's SDK markers, an observed precision of 0.13. The errors are concentrated in one account: all 169 IPNinja flags come from a single publisher of templated applications (\texttt{com.artygeekapps.app}$N$), none of whose current builds, including its three corpus packages, contains the IPNinja SDK. The classifier recognizes the publisher's template, which it learned from builds that once carried the SDK, not the SDK itself. Excluding that publisher, 26 of the remaining 28 flags are confirmed (precision 0.93), and the model accepts all 26 applications in those catalogs that carry their developer's provider markers. Among the 357 analyzed catalog applications the model does not flag, 3 carry any proxy SDK marker (0.8\%; 95\% upper bound 2.4\%), a false-omission proportion conditional on marker recall, which is below one for Luminati and Oxylabs. The separation within catalogs argues against the simplest form of confounding, a classifier that labels every application of a known publisher, but the template case shows that a shared code base can stand in for the SDK where a publisher's applications are near-copies. What the classifier provides on a developer's wider catalog is therefore a screen whose precision depends on the publisher: high where the publisher's applications differ and the SDK is what they share, and near zero where a template carries the signal. Developer concentration is accordingly evidence that distribution is organised through a small number of publishing accounts, and a prioritisation signal, not a basis for labelling individual applications.

\begin{table}[htbp]
    \centering
    \small
    \setlength{\tabcolsep}{3pt}
    \begin{tabular}{llrrr}
    \toprule
    Developer & Proxy & Corpus & Catalog & Confirmed \\
    \midrule
    Creative Simulators Studio & Luminati & 11 & 1/12 & 1 \\
    My Pocket Animals Studio & Luminati & 8 & 2/8 & 2 \\
    Animals Wildlife Studio & Luminati & 7 & 1/27 & 1 \\
    Pocket Games Tycoon & Luminati & 7 & 0/17 & 0 \\
    Mobile Sim Universe & Luminati & 6 & 1/18 & 1 \\
    CrazyFun Games & Luminati & 5 & 2/11 & 2 \\
    FSM SOFT & Luminati & 5 & 0/7 & 0 \\
    No Limit Action & Luminati & 5 & 2/14 & 2 \\
    Power Gaming & Luminati & 5 & 2/33 & 2 \\
    Captaindroid & CashRaven & 4 & 0/14 & 0 \\
    Craft World Team & Luminati & 4 & 2/10 & 2 \\
    Epic Battle Studio & Luminati & 3 & 2/20 & 2 \\
    geekApps & IPNinja & 3 & 169/300 & 0 \\
    Mobile Simulators World & Luminati & 3 & 6/29 & 6 \\
    Assassin Revenge & Luminati & 2 & 0/22 & 0 \\
    Colorful Lab & Luminati & 2 & 0/9 & 0 \\
    Epic mobile studio & Luminati & 2 & 0/19 & 0 \\
    Fury Games Team & Luminati & 2 & 1/23 & 1 \\
    Gaming Fever & Luminati & 2 & 0/19 & 0 \\
    Pocket Sports Life & Luminati & 2 & 4/19 & 4 \\
    RUD Present & Popa & 2 & 1/10 & 0 \\
    Super Sport Team & Luminati & 2 & 0/15 & 0 \\
    Synotic Apps LLC. & Monkeysocks & 2 & 1/6 & 0 \\
    \bottomrule
    \end{tabular}
    \caption{The 23 developer accounts with at least two corpus packages whose current APKPure versions retain their original provider label. ``Corpus'': such packages. ``Catalog'': applications outside the corpus assigned to the listed provider, of those analyzed from the developer's APKPure catalog. ``Confirmed'': flagged applications carrying the provider's SDK markers.}
    \label{tab:retained_devs}
\end{table}

\subsection{Robustness of the pipeline}\label{sec:robustness}
The classifier pipeline operates on a fixed-dimensional WL vector derived from the CFG and FCG, which decouples attribution from the raw bytecode and shapes how the pipeline responds to common transformations.

Identifier renaming is not absorbed by WL hashing when method and class names contribute to the initial node labels. Renaming developer identifiers per application changes 70\% of the predictions of an \textsc{identifier} model trained on unmodified applications (Section~\ref{sec:res_limits}). The \textsc{combined} coloring, built from fuzzy type descriptors, framework-API sets and opcode multisets, is invariant to the transformation by construction, at a cost of 4 points of macro-F1 under DEX grouping and 7 under lineage grouping. Dead-code insertion that leaves CFG topology unchanged, basic-block reordering, or repackaging of the host application may leave portions of the graph unchanged, but robustness cannot be assumed when a transformation changes initial node labels.

Topological transformations are more impactful against this type of pipeline. Aggressive control-flow flattening that routes direct branches through a single dispatch hub, or runtime class loading that removes edges from the static FCG, would alter the vector regardless of the downstream classifier. Further, this information cannot be recovered without dynamic analysis. This is a fundamental limitation of static representations of code.

Unreachable code is a related concern, since androguard's CFG and FCG include methods and basic blocks that are never executed at runtime, which could in principle inject spurious features into the WL vector. Two properties of the pipeline mitigate this. First, the graph-extraction filtering step (Section~\ref{sec:graph_extraction}) removes standard library and common third-party packages that account for much of the unreachable-code surface. Second, WL aggregates structural context from neighboring nodes, so isolated unreachable methods with no callers or callees contribute a small number of distinctive hash buckets and are unlikely to surface as top discriminative features. Targeted padding designed to shift specific WL hash buckets would require the adversary to know the hash dimension $d$ and the hash function in use, although untargeted padding could still shift the bucket histogram without that knowledge; the present evaluation does not measure robustness against adversarial padding.

\subsection{Limitations}\label{sec:limitations}
Our dataset is limited to applications identified through the Mi et al.\ corpus, DEX overlap enumeration, the version histories of labeled packages, and applications reported as contacting published command-and-control domains. There are certainly additional residential proxy applications in the wild, but there is currently no reliable method for quickly enumerating them beyond the code reuse and infrastructure relationships we have already explored. The dataset is also limited to eight proxy families, four of them added from command-and-control domains with few applications, and a detector trained without a provider rarely finds it (Table~\ref{tab:lofo}), so these techniques may not generalize to emerging providers. Additionally, our analysis is purely static; the obfuscation and unreachable-code considerations in Section~\ref{sec:robustness} apply. The trained \texttt{non\_proxy} class is a finite sample of applications rather than a representative model of all unknown software, so deployment requires a calibrated rejection threshold (Section~\ref{sec:opensetproto}). Finally, the ecosystem analysis inherits the limits of its verification: marker recall is below one for Luminati and Oxylabs, 80\% of corpus packages are no longer listed on APKPure, and availability and successful static analysis introduce further selection effects. Future work with additional families would strengthen the generalizability of these results.
\section{Summary and Future Works}\label{sec:conclusion}
Mobile applications that serve as exit nodes for residential proxy networks occupy a dual role, providing a service to the proxy provider while exposing the device owner to risk from third-party traffic routed through their connection. Understanding the ecosystem around these applications has implications both for mobile device owners and for the broader security community's ability to attribute traffic to specific proxy networks.

We find that machine learning classifiers trained on Weisfeiler-Lehman graph kernel features identify residential proxy family membership at macro-F1 = 0.950 over eight providers under DEX-grouped cross-validation, on par with code-only Drebin features, and at 0.722 when applications of the same host lineage are kept in one fold; the same representation extends to a nine-class open-world setting (eight proxy families plus a non-proxy class) at macro-F1 = 0.941. Against 4,887 independently collected Google Play applications, maximum-confidence rejection reaches an AUROC of 0.982, but at 95\% coverage of known applications it still assigns a provider to 5.2\% of Play applications. Applied to current APKPure versions and developer catalogs, the trained classifier finds that 80\% of corpus packages are no longer listed and that 65.6\% of those still analyzable retain their original proxy family, 93\% of them confirmed by the SDK's markers, and it surfaces 23 developer accounts with several retained packages. On their wider catalogs it is precise except where a publisher's template carries the signal, so prospective detection on previously unlabeled inputs is a triage screen rather than a basis for labeling individual applications.

\subsection{Future Work}
As a next step, we will develop a dataset of labeled residential proxy traffic consisting of both benign and malicious flows. Our goal is to combine dynamic analysis techniques with the static analysis presented in this paper to gain a more complete understanding of how these applications interact with the network stack. This would enable detection techniques that operate at both the application and network layers, allowing defenders to identify proxy traffic even when the application itself is not directly accessible for analysis.

\bibliographystyle{IEEEtran}
\bibliography{citations}
\appendix
\section{Ethics Considerations}\label{sec:ethics}

This work analyzes application binaries that were already publicly distributed. It involves no human subjects, collects no personal data, and did not require interaction with any user of the applications studied. We consider the parties affected by the research procedures and by publication, and apply the Menlo Report principles to each rather than restating the framework.

\paragraph{Device owners.}
Owners of devices running proxy applications are the population placed at risk by the ecosystem we study: their bandwidth and IP reputation are used to carry third-party traffic, frequently without meaningful disclosure. They are the intended beneficiaries of this work. Because our analysis is entirely static, we never executed a proxy SDK, never joined a proxy network, and never relayed any third-party traffic through our own infrastructure or through anyone else's device. This is the principal reason we did not pursue dynamic analysis for the present study.

\paragraph{Application developers.}
Our ecosystem measurement identifies developer accounts whose applications repeatedly carry the same proxy family, and we report those accounts by name. We considered anonymizing them and concluded that naming is both defensible and necessary here. The identities are publisher names as listed publicly on the marketplace, attached by the publishers themselves to software they chose to distribute publicly. The underlying fact we attribute to each account, that it published at least two applications carrying the labelled family's SDK, is ground truth from the corpus rather than a model output. Withholding the names would also make the central ecosystem claim unverifiable by a reader, which conflicts with the open science commitments in Appendix~\ref{sec:openscience}, and would remove precisely the information that makes the finding actionable for marketplace operators.

What we take care over is not the names but the strength of claim attached to them. Table~\ref{tab:retained_devs} reports three quantities of different evidentiary weight, and we keep them distinct. The corpus counts rest on the ground-truth labels of the developer's corpus applications. The catalog counts are classifier predictions over applications for which we have no independent label, and the confirmed counts are those predictions whose analyzed build carries the provider's SDK markers, which establishes the presence of SDK code but not that it is active. We describe unconfirmed catalog predictions only as pivots for further investigation, and we report that one named publisher's flagged applications carry no SDK at all, so that the table cannot be read as a finding that every flagged application contains proxy code. We also make no claim that any developer acted unlawfully: embedding a proxy SDK is in many cases a disclosed monetization choice, and some of these accounts may themselves have been supplied an SDK whose behavior was not fully described to them.

\paragraph{Proxy providers.}
The four providers studied are commercial entities, several of which operate openly and describe their consent mechanisms publicly. We make no claim about the legality of any provider's operations. We characterize these applications as potentially unwanted programs on the basis of their observable technical behavior and disclosure practices, not as an allegation of criminal conduct.

\paragraph{Marketplaces and the wider public.}
Marketplace operators and defenders benefit from an attribution method that requires only the application binary. Against this we weigh the dual-use risk that publishing an attribution technique also indicates how to evade it. We judge this risk to be modest: our own results show that identifier renaming, an already-ubiquitous and freely available transformation, substantially degrades label-dependent features, so we disclose no evasion capability that an adversary does not already possess. Reporting this openly, together with the obfuscation-invariant configuration that resists it, is more useful to defenders than withholding it would be.

\paragraph{Data sources and the research team.}
Samples were obtained through AndroZoo's research access programme, the VirusTotal academic API, and public download endpoints, in each case under the terms of use of the provider and with request rates limited so as not to burden the services. We did not circumvent any access control, and we redistribute no sample in a way those terms prohibit (Appendix~\ref{sec:openscience}). Potentially malicious binaries were handled in an isolated analysis environment and were never executed.

\paragraph{Institutional review.}
The study does not constitute human subjects research and accordingly did not require review by our institution's IRB.

\section{Open Science}\label{sec:openscience}

We release the code, derived data, and result files needed to reproduce every quantitative claim in this paper. Artifacts are available at \url{https://anonymous.4open.science/r/static_artifacts_eurosp-C707/}.

\paragraph{What is released.} The repository contains the corpus manifest, giving for each build its SHA-256, provider label, label source, validation status, SDK-marker confirmation and DEX and lineage group, together with the Weisfeiler-Lehman identifier and combined vectors of every evaluated application for both graph types and the MaMaDroid family features (23\,MB), which retrain the classification baselines without any application binaries. It also contains the machine-readable result file behind every reported number, including the per-package records of the ecosystem measurement, a script that recomputes every number printed in this paper from those files and prints each command with its result, for comparison with the paper, and the experiment scripts as run. Scripts requiring credentials we do not distribute, namely the marketplace download and lookup paths, are deliberately omitted, and no APK or DEX file is included. The bundle is 27\,MB.

\paragraph{What cannot be released, and why.}
We do not redistribute the application binaries. The corpus and its derived graph files exceed 450\,GB, which is beyond what any anonymous or archival host will accept, and independently the APKs are third-party copyrighted software obtained under terms that permit research use but not redistribution. We therefore release the SHA-256 manifest of every sample used, together with its family label, source, and DEX group assignment. Any researcher granted AndroZoo access can reconstitute the corpus exactly from this manifest, which is the established practice for Android measurement work of this kind \cite{allix2016androzoo}.

\paragraph{Representative samples.}
Because the complete corpus cannot be shipped, the repository includes a stratified subsample spanning all four families and both the original and expanded datasets, with precomputed call graphs and feature vectors. This subsample is sized to run on a commodity machine and is sufficient to execute the pipeline end to end, so that the extraction and vectorization stages can be verified directly rather than taken on trust. The headline classification and open-set results are additionally reproducible from the released feature vectors and result files without access to any APK.

\paragraph{Reproduction and known limits.}
Results are deterministic given a fixed random seed, which is recorded in the repository and used throughout. Two qualifications apply. The ecosystem measurements of Section~\ref{sec:ecosystem} query a live marketplace and will therefore drift as applications are added and removed; we release the retrieved manifest with collection timestamps so the analysis can be rerun against the recorded snapshot.

The camera-ready version will replace the anonymous links above with a permanent, non-anonymous archival record.

\section{Supplementary Results}\label{sec:appendix}
This appendix contains supporting material referenced from the main text. The paper is self-contained without it.

\subsection{SDK marker verification}\label{sec:app_verification}
Section~\ref{sec:ecosystem} verifies predictions on the builds they were made on: every current build downloaded for classification is scanned for the providers' SDK markers, the namespaces, class names, protocol strings and native library names recovered by reverse engineering, and its SHA-256 is recorded with the prediction. A marker miss is evidence of absence only in proportion to how reliably the marker fires. On the validated corpus, where the provider is ground truth, a provider's markers are found in 99.9\% of Monkeysocks, 92\% of Luminati and 19\% of Oxylabs applications and in every application of the other five providers (Table~\ref{tab:dataset}); Oxylabs chooses its namespace per publisher, so a missing Oxylabs marker carries little information, and only two current Oxylabs builds were analyzable. Marker matching is also the strongest available string-based comparison, benchmarking hand-derived markers rather than automatically generated signatures: it recovers between 0.19 and 1.00 of known applications depending on the provider, while the structural classifier attains 0.950 macro-F1 across all eight.

\end{document}